\documentclass[runningheads]{llncs}
\usepackage{graphicx}
\usepackage{amsfonts}
\usepackage{dirtytalk}
\usepackage{xcolor}
\usepackage{float}
\usepackage[hidelinks]{hyperref}

\makeatletter
\newcommand{\printfnsymbol}[1]{%
  \textsuperscript{\@fnsymbol{#1}}%
}
\makeatother

\begin{document}
\title{Uncertainty-Based Dynamic Graph Neighborhoods For Medical Segmentation}
\titlerunning{Uncertainty-Based Dynamic Graph Neighborhoods}
%

\author{
    Ufuk Demir\thanks{The authors have equal contribution. \newline
    This work is accepted for publication in the PRedictive Intelligence in
MEdicine (PRIME) workshop Springer proceedings in conjunction with MICCAI
2021.} \inst{1}\href{https://orcid.org/0000-0002-5904-146X}{\includegraphics[keepaspectratio=true,scale=0.06]{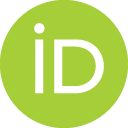}} \and
    Atahan Ozer\printfnsymbol{1}\inst{1}\href{https://orcid.org/0000-0002-3345-3845}{\includegraphics[keepaspectratio=true,scale=0.06]{ORCIDiD_icon128x128.png}} \and
    Yusuf H. Sahin \inst{1}\href{https://orcid.org/0000-0002-4134-9423}{\includegraphics[keepaspectratio=true,scale=0.06]{ORCIDiD_icon128x128.png}} \and
    Gozde Unal \inst{2}\href{https://orcid.org/0000-0001-5942-8966}{\includegraphics[keepaspectratio=true,scale=0.06]{ORCIDiD_icon128x128.png}}
}

\authorrunning{U. Demir et al.}
%
%

\institute{Istanbul Technical University, Faculty of Computer and Informatics Engineering, Computer Engineering, Istanbul, 34469, Turkey \and
Istanbul Technical University, Faculty of Computer and Informatics Engineering, AI and Data Engineering, Istanbul, 34469, Turkey \\
\email{\{demiruf17,ozera17,sahinyu,gozde.unal\}@itu.edu.tr} \\
\url{https://vision.itu.edu.tr}
}

\maketitle              
\begin{abstract}
In recent years, deep learning based methods have shown success in essential medical image analysis tasks such as segmentation. Post-processing and refining the results of segmentation is a common practice to decrease the misclassifications originating from the segmentation network. In addition to widely used methods like Conditional Random Fields (CRFs) which focus on the structure of the segmented volume/area, a graph-based recent approach makes use of certain and uncertain points in a graph and refines the segmentation according to a small graph convolutional network (GCN). However, there are two drawbacks of the approach: most of the edges in the graph are assigned randomly and the GCN is trained independently from the segmentation network. To address these issues, we define a new neighbor-selection mechanism according to feature distances and combine the two networks in the training procedure. According to the experimental results on pancreas segmentation from Computed Tomography (CT) images, we demonstrate improvement in the quantitative measures. Also, examining the dynamic neighbors created by our method, edges between semantically similar image parts are observed. The proposed method also shows qualitative enhancements in the segmentation maps, as demonstrated in the visual results.

\end{abstract}

\section{Introduction}

Deep convolutional neural networks (CNN) have proven to be powerful for computer vision tasks including classification, segmentation, and retrieval \cite{anwar2018medical}. Considering the time spent on the manual segmentation by the medical experts for making quantitative measurements, medical image segmentation by CNNs attained wide usage especially in recent years \cite{ronneberger2015u,milletari2016v,milletari2017hough,kamnitsas2017efficient}. However, despite the fact that many of these models create a rough and useful segmentation, especially for segmented organ borders or organs with similar tissue (e.g. pancreas parenchyma has a similar contrast with the bowel in CT imaging \cite{boers2020interactive}), they can produce unreliable results. Those regions can be presumed by their pixel-wise uncertainty values at the test time. Considering the Monte Carlo Drop Out \cite{Kendal_Gal} (MCDO) uncertainty of these areas in the segmentation network output, Ding et al. developed an uncertainty-aware training procedure that focuses on segmenting the relatively certain parts correctly and reserving the remaining uncertain parts for expert decision \cite{ding2020uncertainty}. Although their method improved the segmentation for certain areas, it is still challenging to manually segment the uncertain regions.

To improve the CNN's results in medical image segmentation, post-processing methods like conditional random fields (CRF) \cite{krahenbuhl2011efficient} could be directly applied to improve the network predictions \cite{kamnitsas2017efficient,feng2020study}. However, the CRF depends strongly on the shape priors and is an independent process from the network features. In \cite{dias2018semantic}, a region growing algorithm to refine the network output focusing on the uncertain pixels is used and better results than a CRF process are obtained. Similarly, to benefit from the network uncertainties in the post-processing step, Soberanis-Mukul et al. developed a graph convolutional network (GCN) Refinement technique \cite{our_baseline}. The main contributions of this work focus on two advancements over the GCN Refinement procedure as follows:

\begin{itemize}
    \item A new dynamic neighbor selection mechanism is defined.
    \item The dynamic neighbor selection mechanism is applied in two setups: inter-graph and intra-graph. 
\end{itemize}

Then, these methods on the neighbor selection are applied for an uncertainty-aware training strategy in which the GCN and the segmentation network are trained end-to-end. Our method increases the performance of the segmentation in a selected application, which is the pancreas segmentation from CT images.

\section{Related Work}

\subsubsection{Uncertainty Estimation:}

Uncertainty estimation is a critical task in automated medical imaging. In general, uncertainty can be modeled in two ways, aleatoric and epistemic uncertainty \cite{uncertainity_epis_aleo}. Aleatoric uncertainty is related to noisy observations of the distribution. On the other hand, epistemic uncertainty is related to deficient observations of the distribution \cite{Kendal_Gal}. The second one can be reduced given enough data however, the first one can not be reduced without removing the source of the noise. Kendal and Gal \cite{Kendal_Gal} proposed it is possible to estimate epistemic uncertainty by using drop-out layers of the model and named this process as Monte Carlo Dropout (MCDO). This metric can provide uncertainty estimation per pixel during segmentation and hence it is possible to find possible erroneous predictions using pixels with high uncertainty.

\subsubsection{Graph Neural Networks:}

The prominence of Graph Neural Networks (GNNs) is increasing due to the latest advancements in the area \cite{graph_attention,pna}.  One of the works that accelerated the research is Graph Convolutional Neural Networks (GCN) by Kipf and Weil \cite{Kipf}. They presented a convolution-based layer propagation network that can directly work in a graph structure. The capabilities of various GNN \cite{Aggregated_CNN_GCN,GRAPHSAGE} models in terms of aggregation schemes is investigated by Xu et al. \cite{xu} in a mathematical frame to characterize the expressive power of GNNs. It is shown that single aggregators may fail to distinguish representations for node classification. The recent work of Corso et al. \cite{pna} shows multiple aggregation functions in GNNs are required to maximize information extraction from the network and presented Principle Neighborhood Aggregation (PNA) blocks.

\subsubsection{Dynamic Graph Neural Networks:} For point cloud segmentation and classification tasks with graph neural networks, nearest spatial neighbors according to point coordinates could be selected to create the edges between the points \cite{zhang2018graph}. However, in Dynamic Graph CNNs (DG-CNN) \cite{wang2019dynamic},  after every network layer, different edges are created according to feature distances between each point. Thus, at the beginning of the network, edges indicate the spatially close points, whereas at the end, they indicate semantically close points.

\paragraph{}
Our work utilizes uncertainty-aware CNN training by using graphs that are constructed by the MCDO process during segmentation network training. Uncertainty graphs are refined by employing multiple aggregator functions and dynamic edge calculations.

\section{Method}
\subsection{GCN Refinement}

We take the GCN Refinement work of Soberanis-Mukul et al. \cite{our_baseline} as our baseline. In GCN Refinement, a graph is created by selecting the uncertain voxels and some certain voxels next to the uncertain ones as graph nodes. Then, for every node, 6 edges are created according to a 6-neighborhood, and 16 edges are created randomly. Using the prediction outputs of certain nodes with low uncertainty, a GCN model is trained in a semi-supervised manner.

In this paper, we focus on two main issues of the GCN Refinement procedure. First, the random neighbor selection process is problematic since using random neighbors, edges could be created between unrelative nodes, and the reproducibility of the technique could be decreased. Second, the network's contribution to the GCN is limited since the training process is not executed in an end-to-end manner. To address those limitations, we devise a dynamic neighbor selection, which is investigated through both an intra-graph and inter-graph edge selection procedure.

Some related definitions that are used throughout the paper are presented next. The uncertainty value $\mathbb{U}(x)$ for the voxel at coordinate $x$ of a segmentation output is calculated using the entropy,
    
\begin{equation}
\mathbb{U}(x)=-\sum_{c=1}^{M} P(x)^{c} \log P(x)^{c},
\end{equation}

\noindent where $P(x)^{c}$ is the probability of the voxel belonging to the class $c$. This probability is estimated by using the expectation of the MCDO process with T passes as

\begin{equation}
\mathbb{E}(x)=\frac{1}{T}\sum_{t=1}^{T} g(V(x), \theta_t),
\end{equation}
 
\noindent where $g$ represents the segmentation network, $V(x)$ represents the voxel intensity and $\theta_t$ represents the network parameters. For each node of the graph, expectation $\mathbb{E}(x)$, entropy $\mathbb{U}(x)$, and voxel value $V(x)$ are used as node features. Weights for the edges are calculated as weighted summation of three different metrics such as expectation diversity $div(x_i, x_j)$ \cite{zhou2017fine} (Eq.\ref{eq:3}), relative intensity $int(x_i, x_j)$ (Eq.\ref{eq:4}) and relative 3-D position $pos(x_i, x_j)$ (Eq.\ref{eq:4}), as follows: 
\begin{equation}
    \label{eq:3}
    div(x_i, x_j) = \sum_{c=1}^{M}\left(P^{c}\left(x_{i}\right)-P^{c}\left(x_{j}\right)\right) \log \frac{P^{c}\left(x_{i}\right)}{P^{c}\left(x_{j}\right)},
\end{equation}

\begin{equation}
    \label{eq:4}
    int(x_i, x_j) = \exp \left(-\frac{\left\|V(x)-V\left(x_{j}\right)\right\|^{2}}{2 \sigma_{1}}\right),
\end{equation}

\begin{equation}
    \label{eq:5}
  pos(x_i, x_j) = \exp \left(-\frac{\left\|x_{i}-x_{j}\right\|^{2}}{2 \sigma_{2}}\right).
\end{equation}

In our work, we keep uncertainty calculation and graph features the same as the baseline work. Our novelty lies in our connectivity structure and uncertainty-aware CNN training as described in the next section.

\subsection{Dynamic Edge Selection and Uncertainty Aware Training}

\begin{figure}
    \centering
    \includegraphics[width=\textwidth]{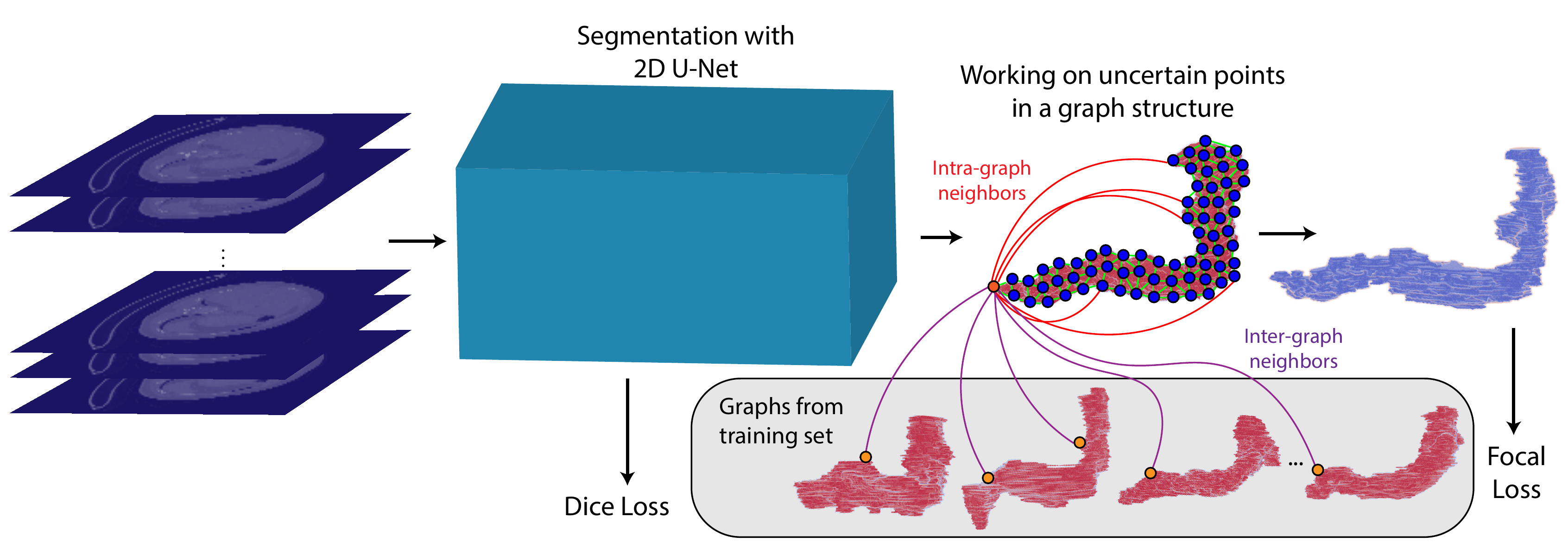}
    \caption{In our uncertainty-aware training procedure, the GCN is trained combined with the segmentation network. In the GCN, in addition to the edges created using 6-neighborhood, intra-graph neighbors or inter-graph neighbors are also selected. Neighbors for only one node are shown in the figure.}
    \label{fig:unc_aware_model}
\end{figure}

The input CT image slices are fed to a 2D segmentation network, which is a standard U-Net \cite{ronneberger2015u} model. The network is trained until it converges, then it continues training with the graph-based method. At this stage, for each iteration, the uncertainty analysis is performed on the input volume. Then, a graph model for the volume, its edges, and graph features are created to use in the GCN. Figure \ref{fig:unc_aware_model} illustrates the proposed method.

In \cite{our_baseline}, two types of edge creation mechanisms are used. First, for every node, connections are created to the 6-neighbors. These local connections bring regional information about each node, however, as the neighbor nodes have nearly the same features with a selected node, the global graph topology is not discovered until the last layers of the network. Second, in order to tolerate the aforementioned problem, a set of randomly selected 16 neighbors are added to each node, and the corresponding edges are created. The random selection process helps to improve the quantitative results, however, it lacks the reproducibility of the results. Our hypothesis is that instead of choosing random neighbors, neighbors that are chosen considering feature distances could improve the quantitative results and their interpretability. Thus, inspired by the neighbor-creation mechanism in DG-CNN, we define two different types of neighbors in the feature space: intra-graph and inter-graph neighbors.

\textbf{\textit{Intra-graph Neighbors: }} As represented by the red lines in Figure \ref {fig:unc_aware_model}, after the patient graph is constructed, the k-nearest neighbors algorithm is applied to select the nearest 5 new nodes from the same patient graph for each node.

\textbf{\textit{Inter-graph Neighbors: }} For each training sample used to train the segmentation network, graphs are created and graph features are calculated individually. Then, for each node of the test graph, a total of 5 new neighbors are selected as illustrated by the purple lines in Figure \ref{fig:unc_aware_model}, according to dynamic feature distances obtained from graphs of training samples. To decrease the memory need, farthest point sampling \cite{qi2017pointnet++} with ratio $\frac{1}{40}$ is applied on the train graphs. 

Both inter-graph and intra-graph neighbor selection mechanisms can be applied to two different procedures: refinement and uncertainty-aware training (UAT)\footnote{Our usage of the term Uncertainty-Aware Training is different from Ding et al. \cite{ding2020uncertainty}.}.

\textbf{\textit{Refinement: }} The refinement procedure is similar to the one presented in GCN Refinement \cite{our_baseline}. All voxels under a segmentation mask of the test object are refined in a semi-supervised manner. The segmentation network is not affected by this procedure.

\textbf{\textit{Uncertainty-Aware Training (UAT): }} We combined the graph network with the segmentation network and trained using the training set in a supervised manner. Thus wrongly labeled voxels have the chance to be corrected by the graph during the train time. For each backward passing, we applied the losses only for one slice for the segmentation network to decrease the required memory constraint.

\begin{figure}
    \centering
    \includegraphics[width=0.8\textwidth]{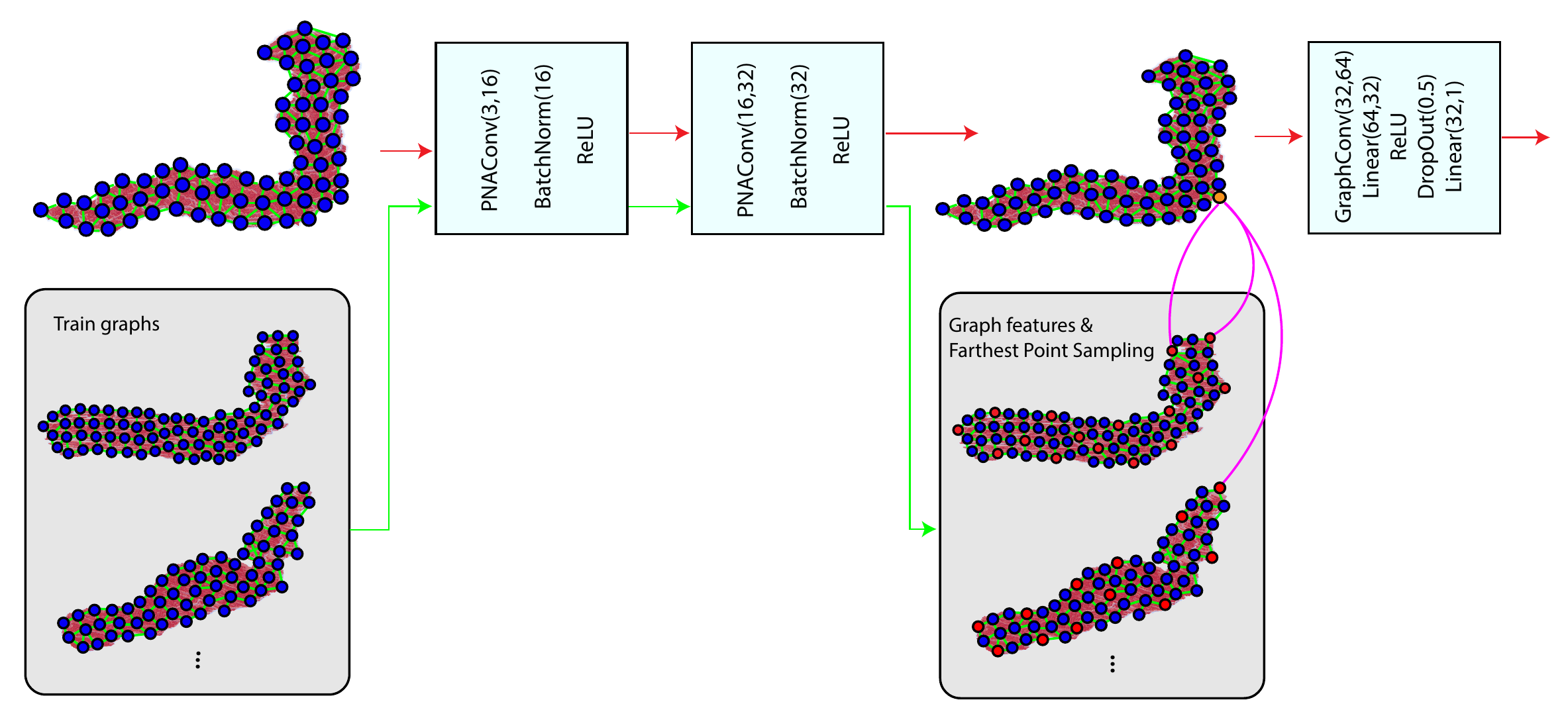}
    \caption{Graph network model used for the "Inter-graph Neighbors" setup. Graph features are calculated in the PNA blocks for both the graph to refine and the train graphs. Then, new edges are created using graph features.}
    \label{fig:graph_model}
\end{figure}

For all studies, a simple network containing two PNA blocks and a GCN layer is used. For the PNA blocks, \textit{mean, min, max, std} aggregators; and \textit{identity, amplification, attenuation} scalers are used. The graph model and its usage for inter-graph neighbors are as given in Figure \ref{fig:graph_model}.

\section{Experiments}

For a fair evaluation, the same pancreas CT dataset from NIH \cite{dataset} and the U-Net \cite{ronneberger2015u} segmentation model officially shared for the GCN Refinement are used. Considering the hyper-parameters and training procedure from the baseline, the GCN part is trained using an Adam \cite{kingma2014adam} optimizer with a learning rate $1e^{-2}$. To ensure the balance between the GCN part and the U-Net, a learning rate of the U-Net is selected as $1e^{-5}$. The number of nodes and edges are selected heuristically. The other hyper-parameters are selected as the same as the baseline. The U-Net model is trained alone for more than 50 epochs until its performance converged. For the U-Net model and GCN, dice loss and focal loss are used respectively. For all experiments, we used the PyTorch 1.7.1 framework  and PyTorch Geometric \cite{fey2019fast} library. We trained the models on a device having Titan RTX.

To evaluate our method, we designed the experiments given in Table \ref{experiments}.
We execute 4 modes of our method: our own base methods \say{Intra-graph} and \say{Inter-graph}; as well as the \say{Intra-graph UAT} and \say{Inter-graph UAT} with the uncertainty-aware components.\footnote{\url{https://github.com/ituvisionlab/Uncertainty-Based-Dynamic-Graph-Neighborhoods}}

\begin{table}
\centering
\begin{tabular}{|l|cccc|}
\hline
Experiment      & 6-Neighbors & \begin{tabular}[c]{@{}l@{}}Intra-Graph\\ Neighbors\end{tabular} & \begin{tabular}[c]{@{}l@{}}Inter-Graph\\ Neighbors\end{tabular} & \multicolumn{1}{c|}{\begin{tabular}[c]{@{}c@{}}Uncertainty-Aware\\ Training\end{tabular}} \\ \hline
6-Connectivity  & \checkmark           &                                                                 &                                                                 &                                                                                           \\
Intra-graph     & \checkmark           & \checkmark                                                               &                                                                 &                                                                                           \\
Inter-graph     & \checkmark           &                                                                 & \checkmark                                                               &                                                                                           \\
Intra-graph UAT & \checkmark           & \checkmark                                                               &                                                                 & \checkmark                                                                                         \\
Inter-graph UAT & \checkmark           &                                                                 & \checkmark                                                               & \checkmark                                                                                         \\ \hline
\hline
\end{tabular}
\caption{The list of experiments to evaluate our method on pancreas segmentation on CT images.}
\label{experiments}
\end{table}

\subsection{Quantitative Results}

For both refinement and UAT setups, the Dice scores for the test set are given in Table \ref{table:dices1} and \ref{table:dices2} respectively. In both setups, our method overperforms the baseline GCN refinement. Detailed explanations for each experiment's results are as below.

\textbf{6-Connectivity:} Since connectivity is dramatically reduced, the GCN model suffered from deficient information. In fact, it decreases the segmentation network's performance. We can interpret that even random neighbors are useful to keep the contextual information as in the original GCN Refinement.

\textbf{Inter-graph:} Since connectivity is increased quantitatively and semantically meaningful edges are created, our method overperformed the baseline. 

\textbf{Intra-graph:} Using the connections inside the same patient's graph, better results than  Inter-graph refinement are obtained. The best refinement performance in terms of the Dice score is achieved with this setup.

\textbf{Inter-graph UAT:} Improved results are obtained compared to the Inter and Intra-graph models in terms of Dice scores and their standard deviations.  These improvements show that our UAT procedure is quantitatively better than refinement.

\textbf{Intra-graph UAT:} This experiment yielded the best scores among all setups. Our deductions for the \textit{Intra-graph} experiment is also valid for this experiment.

\begin{table}
\begin{minipage}{0.48\linewidth}
\begin{tabular}{c|c|}
\cline{2-2}
                                                & Dice Score               \\ \cline{1-2}
\multicolumn{1}{|l|}{GCN Refinement (Baseline)} & $77.81\pm 6.3$             \\ \cline{1-2}
\multicolumn{1}{|l|}{6-Connectivity}            & $76.11 \pm 7.81$           \\ \cline{1-2}
\multicolumn{1}{|l|}{Inter-Graph}               & $78.32 \pm 6.41$            \\ \cline{1-2}
\multicolumn{1}{|l|}{Intra-Graph}               & $\mathbf{78.87 \pm 6.24}$  \\ \cline{1-2}
\hline
\end{tabular}
\caption{Dice score results for Uncertainty Aware Training results.}
\label{table:dices1}
\end{minipage}\hfill
\begin{minipage}{0.48\linewidth}
\begin{tabular}{c|c|}
\cline{2-2}
                                                & Dice Score             \\ \cline{1-2}
\multicolumn{1}{|l|}{GCN Refinement (Baseline)} & $76.9\pm6.6$            \\ \cline{1-2}
\multicolumn{1}{|l|}{Inter-Graph UAT}           & $78.84 \pm 5.84$           \\ \cline{1-2}
\multicolumn{1}{|l|}{Intra-Graph UAT}           & $\mathbf{79.26 \pm 5.78}$  \\ \cline{1-2}
\hline
\end{tabular}
\caption{Dice score results for refinement setups.}
\label{table:dices2}
\end{minipage}
\end{table}

\begin{figure}
    \centering
    \includegraphics[width=\textwidth]{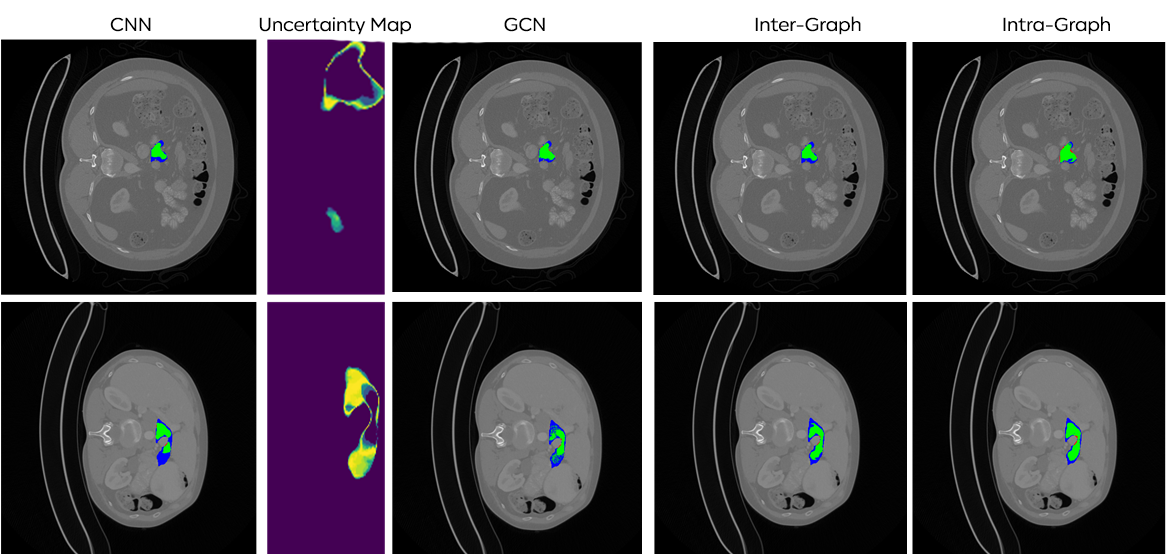}
    \caption{Comparison of the GCN refinement. Red represents false positive, green represents true positive, and blue represent false negative regions. Each row corresponds to another 2D slice. The columns correspond to the CNN model results, the uncertainty map, the GCN Refinement results, the Inter-Graph and the Intra-Graph results respectively.}
    \label{fig:panc_vis_ref}
\end{figure}

\begin{figure}
    \centering
    \includegraphics[width=\textwidth]{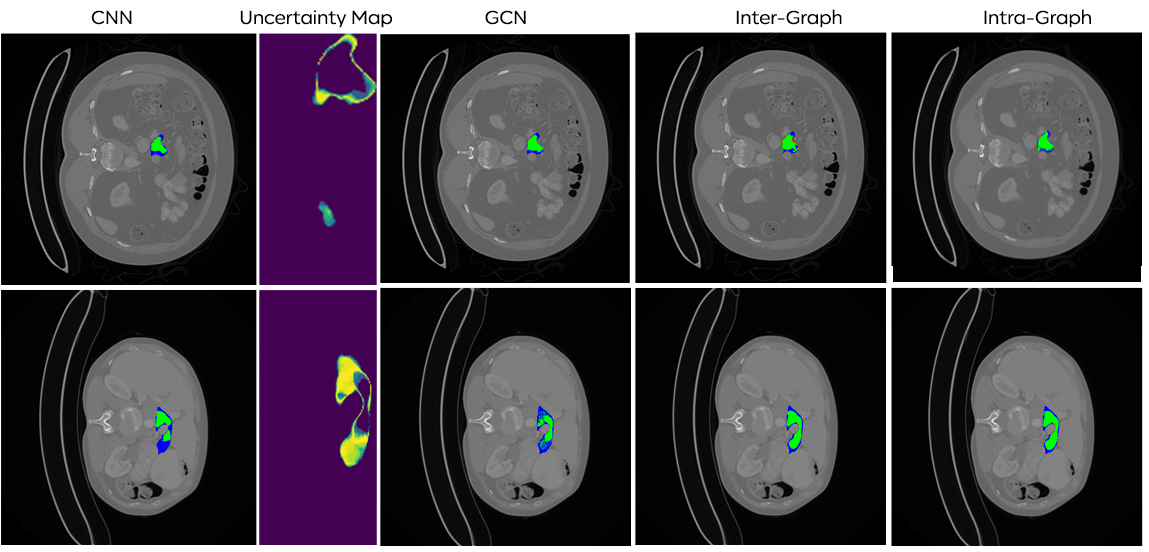}
    \caption{ Visualization of the UAT method.  Same coloring and row order with figure \ref{fig:panc_vis_ref} is applied.   The columns correspond to the CNN model results, the uncertainty map, the GCN Refinement results, the Inter-Graph  with UAT and the Intra-Graph with UAT results respectively.}
    \label{fig:panc_vis_uat}
\end{figure}

In Figure \ref{fig:panc_vis_ref}, visual results of our refinement method are compared with baseline results. To demonstrate the correctly refined parts in uncertain regions, uncertainty maps are also included. As stated in the previous works \cite{ding2020uncertainty,our_baseline}, uncertain regions frequently occur in the border areas. As it can be seen from the visualizations, the refinement method improved the results of uncertain regions both for inter-graph and intra-graph neighborhoods compared to baseline work. The same results are also obtained for the UAT method in Figure \ref{fig:panc_vis_uat}. In UAT visualizations, We observed the best visual results coherently with Dice scores given in Table \ref{table:dices1} and \ref{table:dices2}.

\subsection{Neighboring Results}

\begin{figure}
    \centering
    \includegraphics[width=\textwidth]{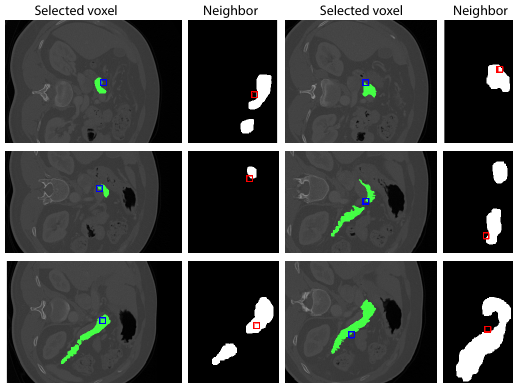}
    \caption{First and third columns show some voxels from pancreas slices, second and fourth columns show pancreas label maps of one of their neighbors. }
    \label{fig:komsu}
\end{figure}

The adequateness of our neighbor selection mechanism can be investigated by checking the selected neighbors. In Figure \ref{fig:komsu}, some voxels from test slices and label maps of their selected neighbors are shown. According to the results, we can conclude that the selected neighbors demonstrated a semantical similarity as argued by Wang et al. for the DG-CNN \cite{wang2019dynamic}. Also for the roughly certain voxels like the ones at the center of the image, the found neighbors are at the more certain positions of the pancreas.

\section{Conclusion}

In this study, we introduced an uncertainty-aware CNN model training procedure and a dynamic edge selection method. Although our graph generation relies on that of the baseline GCN Refinement method, unlike the baseline, we utilize the graph network also for CNN training, allowing the model to learn about uncertain regions in the segmentation. Additionally, inspired by the DG-CNN, we implemented two different neighbor selection methods: Intra-graph and Inter-graph. In our best setup using Intra-graph neighbors for Uncertainty-Aware Training, we obtained an increase of ${\small \sim} 1.45\%$ over GCN Refinement. Investigating the quantitative results for both refinement and UAT, we can infer that using these neighbor selection mechanisms and hence providing more contextual information about the pancreas caused a general improvement over the results.

As future work, the proposed method could be extended to multi-organ segmentation while the CNN part segments multiple organs and for each organ, a different GCN is trained. Also, the method could be applied to larger datasets to investigate whether the quality of the inter-graph neighborhood is dependent on the variety of the dataset.\\

\noindent\textbf{Acknowledgement.} This work is supported by the Scientific Research Project Unit (BAP) of Istanbul Technical University, Project Number: MOA-2019-42321.

\bibliographystyle{splncs04.bst}
\bibliography{refs.bib}

\end{document}